\documentclass[prl,twocolumn,tightenlines,footinbib,superscriptaddress]{revtex4-2}
\usepackage{bm,dcolumn,amsmath,graphicx,amsfonts,amssymb}

\usepackage{hyperref}
\usepackage{xcolor}
\definecolor{cite}{rgb}{0.,0.,0.9} 
\hypersetup{colorlinks,linkcolor={cite},citecolor={cite},urlcolor={cite}}

\usepackage[capitalise]{cleveref}
\usepackage{siunitx}

\usepackage{feynmp-auto}

\renewcommand{\vec}[1]{\ensuremath{{\boldsymbol{#1}}}}
\newcommand{\abs}[1]{\ensuremath{\left |#1\right |}}
\newcommand{\be}{\begin{equation}}	
\newcommand{\ee}{\end{equation}}	

\usepackage[normalem]{ulem}
\definecolor{newc}{rgb}{0.,0.6,0.4}

\renewcommand{\section}[1]{\vspace{0.85pt}\paragraph*{\textbf{\textit{\small{#1---}}}}}
\renewcommand{\subsection}[1]{\paragraph*{{\textit{\small{#1---}}}}}

\newcommand{\A}{\ensuremath{\mathcal{A}}} %symbol for hyperfine constant

\usepackage{changes}
\definechangesauthor[name={Natalia}, color=magenta]{no}
\definechangesauthor[name={Jacinda}, color=black!30!green]{jg}
\definechangesauthor[name={Odile}, color=orange]{os}
\definechangesauthor[name={Igor}, color=blue]{iv}

\begin{document}

\title{Smallness of the nuclear polarization effect in the hyperfine structure of heavy muonic atoms as a stimulus for next-generation experiments}

\author{J. Vandeleur}
\affiliation{School of Mathematics and Physics, The University of Queensland, Brisbane QLD 4072, Australia}
\author{G. Sanamyan}
\affiliation{School of Mathematics and Physics, The University of Queensland, Brisbane QLD 4072, Australia}
\author{O. R. Smits}
\affiliation{School of Mathematics and Physics, The University of Queensland, Brisbane QLD 4072, Australia}
\author{I. A. Valuev}
\affiliation{Max Planck Institute for Nuclear Physics, Saupfercheckweg 1, 69117 Heidelberg, Germany}
\author{N. S. Oreshkina}
\affiliation{Max Planck Institute for Nuclear Physics, Saupfercheckweg 1, 69117 Heidelberg, Germany}
\author{J. S. M. Ginges}%\email[]%{j.ginges@uq.edu.au}
\affiliation{School of Mathematics and Physics, The University of Queensland, Brisbane QLD 4072, Australia}
\date{\today}

%-------------------------------------------------------------------------
\begin{abstract}

There is renewed interest in studies of muonic atoms, which may provide detailed information on nuclear structure. A major limiting factor in the interpretation of measurements is the nuclear polarization contribution. 
We propose a method to determine this contribution to the hyperfine structure in muonic atoms from a combination of theory and experiment for hydrogenlike ions and muonic atoms. Applying the method to $^{203,205}$Tl and $^{209}$Bi, for which there are H-like ion and muonic atom hyperfine experimental data, we find that the nuclear polarization contribution for these systems is 
small, and place a limit on its size of less than $10\%$ the total hyperfine splitting. 
We have also performed direct calculations of the nuclear polarization contribution using a semi-analytical model, which indicate that it may be as much as two orders of magnitude smaller. {Therefore, we conclude that the nuclear polarization correction to the hyperfine structure of muonic atoms  does not represent a limiting factor for next-generation experiments.}

\end{abstract}

\maketitle

{\it Introduction.} 
Experiments with muonic atoms are highly sensitive to the effects of nuclear structure, which has been exploited over many decades to provide some of the best information on nuclear charge radii from binding and transition energies, and on nuclear moments and their distribution from  hyperfine splittings~\cite{Wheeler1953,Engfer1974,Borie1982,Buttgenbach1984}. The nuclear structure effects may be orders of magnitude larger than in usual atoms due to the muon's proximity to the nucleus. 
There has been a resurgence of interest
in studies of heavy muonic atoms and what they can offer various areas of nuclear and fundamental physics~\cite{Knecht2020,Valuev2022,Oreshkina2022,Sanamyan2023,Yerokhin2023,Xie2023}, including low-energy precision searches for new physics~\cite{GingesReview2004,SafronovaReview2018}. A new experimental program focused on heavy muonic atoms has begun at the Paul Scherrer Institute~\cite{Knecht2020, PhysRevC.101.054313}.

Despite the remarkable sensitivity to the charge and magnetic nuclear structures in muonic-atom experiments, there are aspects of the theoretical problem that have been anticipated to be sizeable and at the same time considered prohibitively difficult to calculate. Probably the most significant of these is the nuclear polarization contribution~\cite{Plunien1989, Beier2000}. The lack of information on the size of these effects 
compromises the confidence in extracted nuclear properties. 

Progress has been made recently in evaluation of nuclear polarization effects for the fine-structure splitting of heavy muonic atoms~\cite{Valuev2022}. 
The problem is significantly more challenging for the effects in the hyperfine splitting. For light systems, we are aware of direct calculations for hydrogen and muonic hydrogen~\cite{Tomalak2019}. And just recently, the polarization effect in the hyperfine structure of $^3$He$^+$ was derived from a combination of theory and experimental data for magnetic elastic scattering, with the effect found to contribute at about $3\%$ \cite{Patkos2023}, surprisingly smaller than the effect in hydrogen. For heavy systems, there are direct calculations of the nuclear polarization contribution to the hyperfine structure for few-electron highly-charged ions~\cite{Nefiodov2003,Volotka2014}, though we are not aware of evaluation of the effect up until now in heavy muonic atoms. 

In this work, we propose a method to determine the nuclear polarization effect in the hyperfine structure of heavy muonic atoms, by combining experimental data for H-like ions and muonic atoms together with atomic calculations. The nuclear polarization effect may be extracted due to the different sensitivities of the two systems to nuclear structure effects. We apply the method to H-like and muonic $^{203,205}$Tl and $^{209}$Bi, for which experimental data are available. We find that the effect is relatively small for these systems, and we place an upper limit for its value of less than 10\% the size of the total splitting. The small value of the effect is confirmed in the current work from direct calculations using a semi-analytical model, for which substantially smaller values are obtained.  

{\it Semi-empirical approach}. Calculations for both H-like ions and muonic atoms begin with the relativistic treatment of the electron or muon in the Coulomb field of the nucleus, 
\be
    \left[ c \vec{\alpha} \cdot \vec{p} + \left( \beta - 1 \right) m c^2 + V_{\rm nuc} (r) \right] \phi(r)
    = \varepsilon \phi(r) \ ,
    \label{eqn:dirac}
\ee
where $\vec{\alpha}$ and $\beta$ are Dirac matrices, $V_{\rm nuc}(r)$ is the nuclear potential, $m$ is the mass of the electron or muon, $c$ is the speed of light, and $\varepsilon$ is the binding energy. The nuclear potential $V_{\rm nuc}(r)$ corresponds to that of a Fermi charge distribution with root-mean-square charge radii from Ref.~\cite{Angeli2013}. The radial part of the wave functions is defined such that the large and small components, $f(r)$ and $g(r)$, are normalized as $\int_0^\infty \left( f^2 + g^2 \right)dr = 1$. We use atomic units throughout, $\hbar = m_e = \abs{e} = c \alpha = 1$. 

The magnetic hyperfine structure arises from the interaction of the magnetic field produced by the electron or muon with the magnetic moment of the nucleus. This may be described by the relativistic interaction Hamiltonian
\be
    h_{\rm hfs} = \alpha \vec{\mu} \cdot \left( \vec{r} \times \vec{\alpha} \right) F(r)/r^3 \ ,
    \label{eqn:Hhfs}
\ee 
where $\alpha$ is the fine structure constant, $\vec{\mu} = \mu \vec{I}/I$ is the nuclear magnetic moment, $\vec{I}$ is the nuclear spin, and $F(r)$  accounts for the finite distribution of magnetization. For the point-like case $F(r)=1$.

The energy shift caused by the magnetic dipole hyperfine interaction may be calculated in the first order of perturbation theory, and is often expressed in terms of the hyperfine constant $\A$ as:
\begin{equation}
    \Delta E^{\rm hfs} = \langle h_{\rm hfs} \rangle \equiv \frac{F(F+1)-I(I+1)-J(J+1)}{2}\A, 
\end{equation}
where $J$ is the total angular momentum of the lepton and $F$ is the total angular momentum of the system.
The hyperfine constant may be conveniently written in terms of the sum of several distinct contributions,  
\be
\label{eq:aterms}
    \A = \A_0 + \A_{\rm BW} + \A_{\rm QED} + \A_{\rm NP} {+ \A _{\rm h.o.}}\ .
\ee
The first term $\A_0$ is the point-nucleus contribution, which for $s$ states is given by 
\be
    \A_0 = \frac{4}{3} \frac{\alpha}{m_p} \frac{\mu}{\mu_N I} \int_0^\infty dr f(r) g(r) / r^2,
\ee
where $\mu_N$ is the nuclear magneton and $m_p$ the proton mass. This term accounts for the finite nuclear charge distribution by solving the Dirac equation in the Coulomb field of the extended nucleus. The second term is the Bohr-Weisskopf effect $\A_{\rm BW}$, which characterizes the deviation of the radial distribution of the nuclear magnetic moment from the point-like case, evaluated as  
\be
    \A_{\rm BW} = \frac{4}{3} \frac{\alpha}{m_p} \frac{\mu}{\mu_N I} \int_0^\infty dr f(r) g(r) (F(r) - 1) / r^2 \ ,
\ee 
and typically depends on a model for the nuclear magnetization distribution $F(r)$~ 
\cite{Volotka2008, Elizarov2006,Michel2017, Sanamyan2023}. 

The treatment of the quantum electrodynamics (QED) contribution, $\A_{\rm QED}$, is rather different between H-like ions and muonic atoms. For H-like ions, the self-energy (SE) dominates but nevertheless is still comparable with the vacuum polarization~(VP) (see, e.g., \cite{Yerokhin_Lamb_2011}). On the contrary, for muonic atoms, the QED contribution is mostly determined by the VP~\cite{Borie1982,Michel2017}.
This also leads to the change in the sign of the total QED contribution for muonic atoms compared to H-like ions.
In this work, the values for H-like systems for the SE and the dominant VP contributions are from Ref.~\cite{Shabaev1997}, while the higher-order VP contribution is taken from Ref.~\cite{Artemyev2001}. 
For muonic atoms, the leading QED contribution -- the electric and magnetic VP, in the Uehling approximation - has been been evaluated in this work for a uniform magnetization distribution; see Ref.~\cite{Volotka2008} for the expressions. The finite-nucleus corrections to the VP are sizeable, at about $50\%$, and we generously assign a 50\% uncertainty to our values. 

\begin{figure*}
    \centering
    \includegraphics[width=0.8\textwidth]{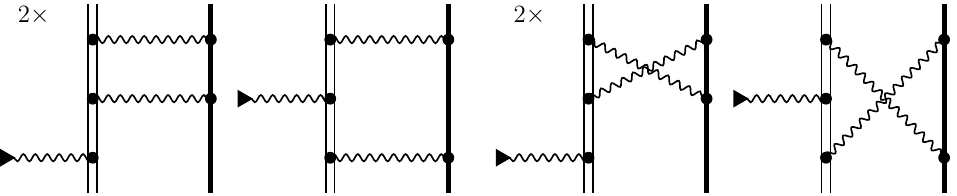}
    \caption{{Leading-order diagrams for the nuclear polarization correction to the hyperfine structure. Here, double line indicates bound lepton, bold line indicates nucleus, wavy lines stand for the photon propagator, and a triangle corresponds to the hyperfine interaction with the magnetic moment of the nucleus. Due to symmetry reasons, the first and third diagrams are to be counted twice.}}
    \label{fig:np}
\end{figure*}

It is also worth mentioning the higher-order contributions $\A_{\rm h.o.}$, which include nuclear deformation, nuclear recoil, and two-loop QED. However, they are estimated to be rather small and completely concealed by the leading terms' uncertainties \cite{Yerokhin_Lamb_2011}. Therefore, we will 
omit them in the following.

Finally, we come to the contribution arising from the polarization of the nucleus by the electron or muon, in the lowest order commonly illustrated by the Feynman diagrams in~\cref{fig:np}. Calculations for hydrogenlike systems show that the nuclear polarization (NP) contribution, $\A_{\rm NP}$, is far smaller than the QED contribution, and within its assigned uncertainty~\cite{Nefiodov2003,Volotka2014}. It may be expected that the relative correction, $\A_{\rm NP}/\A_0$, for muonic atoms is larger than that for hydrogenlike ions, due to the muon's proximity to the nucleus, and the muonic energy spectra being closer to that of the nucleus.

It can be seen that the nuclear polarization contribution may be empirically determined from the measured value of the hyperfine constant $\A_{\rm exp}$, along with accurate account of the other contributions, 
\be
    \A_{\rm NP}^{\rm exp} = \A_{\rm exp} - \A_0 - \A_{\rm QED} - \A_{\rm BW} \ . \label{eq:ANP_exp}
\ee
While $\A_0$ and $\A_{\rm QED}$ may be reliably evaluated for muonic systems, we are faced with the challenge of finding $\A_{\rm BW}$. 
Indeed, modelling of the nuclear magnetization distribution cannot yet be accurately performed from nuclear theory, and it is important to find ways to either eliminate the BW effect or to determine it from some empirical means; see, e.g., Refs.~\cite{Shabaev2001,Elizarov2006,Ginges2018,Skripnikov2020,Roberts2021scr,Sanamyan2023}. 
This is where measured hyperfine data for H-like ions is particularly helpful. 
We can evaluate the BW effect in H-like ions with various nuclear distribution $F(r)$ models, %\cite{Elizarov2006, Sanamyan2023}, 
adjust the parameters of each model such that the experimental value for H-like ions is exactly reproduced, and then apply this fitted model to the muonic atom with the same nucleus.
This follows the same procedure as that in Ref.~\cite{Elizarov2006} and that of our recent work \cite{Sanamyan2023}, where the procedure is performed in reverse. Using the proposed method, the nuclear polarization contribution to the hyperfine structure for heavy muonic atoms may be found.

\begin{table*}[tbh]
    \caption{Nuclear parameters and contributions to hyperfine constants $\A$ for the $1s$ state in H-like ions in eV. The BW values in the final column are determined from the experimental data.     }
    \label{tbl:eHF}
    \begin{ruledtabular}
    \begin{tabular}{ccccccccc}
         &$I^{\pi}$&$\mu\, (\mu_N)$&$r_{\rm rms}$\, (fm) &$\mathcal{A}_{\rm exp}$ & $\mathcal{A}_0$ & $\mathcal{A}_{\rm QED}$\,\cite{Shabaev1997,Artemyev2001} & $\mathcal{A}_{\rm NP}$\,\cite{Nefiodov2003} & $\mathcal{A}_{\rm BW}^{\rm exp}$\\
        \hline
        $^{203}{\rm Tl}^{80+}$ & ${1/2}^+$&1.616(2)&5.4666(27)&3.21351(25)\tablenotemark[1] & 3.293(4) & -0.0188(2) & $0.031(31)\times 10^{-3}$ &-0.061(4)\\ 
        $^{205}{\rm Tl}^{80+}$ & ${1/2}^+$&1.632(2)&5.4759(26)&3.24409(29)\tablenotemark[1] & 3.325(4)& -0.0190(2) & $0.031(31)\times 10^{-3}$ &-0.062(4)\\ 
        %uncertainty of 2 or 3 for QED?
        $^{209}{\rm Bi}^{82+}$ & ${9/2}^-$&4.092(2)&5.5211(26)&1.01701(2)\tablenotemark[2] & 1.0340(5) & -0.00602(5) & $0.011(11)\times 10^{-3}$&-0.0110(5)
        \\
    \end{tabular}
    \end{ruledtabular}
    \tablenotetext[1]{Reference \cite{Beiersdorfer2001};
    $^{\rm b}$\,Reference \cite{Ullmann2017}.}
\end{table*} 

%{\it Results.} 
In~\cref{tbl:eHF} we present the nuclear parameters utilized in our calculations for $^{203,205}$Tl and $^{209}$Bi, including up-to-date recommended values for nuclear magnetic moments, and measured hyperfine constants and calculated contributions for $1s$ hydrogenlike ions. It can be seen that there is a high level of precision for the experimental data $\A_{\rm exp}$, and that the largest uncertainties originate from the theoretical evaluations, in particular corresponding to the dominant contribution $\A_0$. This uncertainty primarily stems from that in the nuclear magnetic moment, a coefficient in $\A_0$. A smaller, and insignificant, contribution to the uncertainty comes from the nuclear rms charge radius uncertainty. 
For the QED contributions, we conservatively assign an uncertainty of 1 in the final digit of the value presented in Ref.~\cite{Shabaev1997}. The NP corrections together with their uncertainties are taken from \cite{Nefiodov2003}.

In the final column of data in \cref{tbl:eHF}, we give the BW correction in the H-like ions, found from the experimental data in combination with the theory contributions. The $F(r)$ reproducing this value is then used to determine $\A_{\rm BW}$ for the muonic atoms.

\begin{table*}[bth]
    \caption{Contributions to the hyperfine constants $\A$ for the $1s$ state in muonic atoms in keV. The two final columns give the nuclear polarization contributions obtained by the semi-empirical and semi-analytical approaches. 
    }    
    \label{tbl:muHF}
    \begin{ruledtabular}
    \begin{tabular}{cccccc c}
      & $\A_{\rm exp}$\cite{Buttgenbach1984} & $\A_0$ & $\A_\text{QED}$ & $\A_{\rm BW}$ & $\A^{\rm {semi-emp}}_{\rm NP}$ & {$\A^{\rm semi-an}_{\rm NP}$}\\
    \hline
    $\mu-^{203}{\rm Tl}^{80+}$ & 2.340(80) & 4.695(6) & 0.037(19) & -2.355(148) & -0.037(169) & 0.0024(24)\\
    $\mu-^{205}{\rm Tl}^{80+}$ & 2.309(35) & 4.726(6) & 0.037(19) & -2.373(149) & -0.081(154) & 0.0024(24) \\ 
    $\mu-^{209}{\rm Bi}^{82+}$ & 0.959(52) & 1.3394(7) & 0.010(5) & -0.412(20) & 0.022(56) & 0.0008(8)\\
    \end{tabular}
    \end{ruledtabular}
\end{table*} 

In \cref{tbl:muHF} we present all contributions to the hyperfine constants for muonic atoms $^{203,205}$Tl and $^{209}$Bi. In the second-to-last column, the empirical value for the nuclear polarization $\A^{\rm semi-emp}_{\rm NP}$ is found from the difference between the experimental value and the contributions $\A_0$, $\A_{\rm QED}$, and $\A_{\rm BW}$ (see \cref{eq:ANP_exp}). 
The uncertainties for the BW effect arise from the
nuclear magnetic moment,  
and the nuclear model dependence in the translation from H-like to muonic systems, in comparable measure. 
The extracted values for the nuclear polarization are dwarfed by the associated uncertainties stemming from the Bohr-Weisskopf effect determination and, for $\mu-$Bi, the uncertainty of the hyperfine measurement. 
Despite these uncertainties, it can be seen that the nuclear polarization contributions are relatively small: less than 10\% the size of the measured hyperfine constants. Compared to the size of the BW effect, they amount to less than 10\% for $\mu-$Tl, and less than about 20\% for $\mu-$Bi.

{\it Semi-analytical model.} 
The second method used for evaluation of the nuclear polarization correction is based on a semi-analytical model, introduced in Ref.~\cite{Plunien1989} and later used in Refs.~\cite{Labzowsky1994,Nefiodov1996, Nefiodov2003,Volotka2014}. 
Recently, this  formalism has been used for calculations of NP corrections to the energies, $g$ factor and hyperfine splitting of H-like ions~\cite{Cakir2020, Schneider_2022}. 

The NP correction to the hyperfine shift of a state $a$, depicted in Fig.~\ref{fig:np}, can be expressed as
\begin{align}
\Delta &E^{\rm NP, hfs}_{a}  = -\alpha \sum\limits_{LM} \sum_{n,k} B(EL) \int\limits_{-\infty}^{\infty} \frac{d\omega}{2\pi i}\frac{2\omega_L}{\omega^2 - \omega_L^2 +i0}  \notag \\
&\times \biggl[ 2\frac{\langle a|F_L Y^*_{LM}|n\rangle \langle n|F_L Y_{LM}|k\rangle\langle k|h_{\rm hfs}|a\rangle}{(\epsilon_a - \omega - \epsilon_n(1-i0))(\epsilon_a  - \epsilon_k(1-i0))}  \\
 &+ \frac{\langle a|F_L Y^*_{LM}|n\rangle \langle n|h_{\rm hfs}|k\rangle \langle k|F_L Y_{LM}|a\rangle}{(\epsilon_a - \omega - \epsilon_n(1-i0))(\epsilon_a  - \omega-\epsilon_k(1-i0))}
\biggr]\,. \notag
\end{align}

The $Y_{LM}$ are spherical harmonics, and the radial functions $F_L$, describing $L$th multipolarity interaction between a lepton and the nucleus, are often taken in their simplest form~\cite{Nefiodov2003} as
\begin{align}
\label{eq:FL}
F_0(r) &= \frac{5\sqrt{\pi}}{2R_0^3}\biggl[1-\biggl(\frac{r}{R_0}\biggr)^2\biggr]\Theta(R_0-r)\,,  \notag \\
F_L(r) &= \frac{4\pi}{(2L+1)R_0^L} \frac{{\rm min}(r,R_0)^L}{{\rm max}(r,R_0)^{L+1}}\,, \qquad L \geq 1\, .
\end{align}  
The first term in the brackets contains both irreducible (for $k,n\neq a$) and reducible ($k,n=a$) contributions, and the factor of 2 in the first term originates from the existence of two equivalent diagrams.
The second term stands for the vertex contribution. 
Although muonic systems are much more sensitive to the details of the radial dependence of $F_L$, the simple \cref{eq:FL} can still be used for an order-of-magnitude estimation of the effect.

The nuclear excitation energies {$\omega_L$} and the reduced transition probabilities $B(EL;L\rightarrow 0)$ have been taken from Refs.~\cite{KONDEV2021509,KONDEV20201,CHEN2015373}.  
For H-like ions, our results obtained by the semi-analytical approach with $F(r)=1$ are in perfect agreement with those from~\cite{Nefiodov2003}, presented in Table~\ref{tbl:eHF}.
The results for the NP corrections to the hyperfine constants for the muonic atoms are listed in the last column of Table~\ref{tbl:muHF}.
In the case of corrections to energy levels, we have found that the simplified semi-analytical model underestimates the NP effect by $\approx 50\%$ compared to recent rigorous calculations \cite{Valuev2022, Valuev2024}.
The results are rather sensitive to the nuclear rms radius and, in the case of hyperfine structure, also to the $F(r)$ function used~\cite{Michel2017, Schneider_2022}. 
Based on this, we conservatively assign to the NP correction the same large uncertainty of 100\% in Table \ref{tbl:muHF} as given in Table \ref{tbl:eHF} for H-like ions.

{\it Discussion and conclusion.}
We have used two independent and complementary approaches to determine the nuclear polarization contribution to the hyperfine structure in muonic atoms. 
We have applied both methods to $^{203,205}$Tl and $^{209}$Bi, and found them to be in agreement with each other. 
Our novel empirical approach, combining hydrogen-like and muonic-atom experiments and theory, gives upper limits on the nuclear polarization contributions of 10\%, or less, the size of the total hyperfine splittings and Bohr-Weisskopf effects. These values can be improved upon in the future with new information about nuclear properties and structure, which dominates the uncertainty. Our semi-analytical approach indicates that the NP contributions are significantly smaller yet, appearing at the level of 0.1\% the size of the BW effect.  
Therefore, we suggest that NP effects can be safely neglected, and the hyperfine structure of muonic atoms may be studied experimentally to improve our knowledge about nuclear magnetization distributions.

\section{Acknowledgments}
We thank J. Hasted for useful discussions. This work was supported by the Australian Government through an Australian Research Council (ARC) Future Fellowship No.\ FT170100452 and ARC Discovery Project DP230101685. 

\bibliography{library}

%apsrev4-2.bst 2019-01-14 (MD) hand-edited version of apsrev4-1.bst
%Control: key (0)
%Control: author (8) initials jnrlst
%Control: editor formatted (1) identically to author
%Control: production of article title (0) allowed
%Control: page (0) single
%Control: year (1) truncated
%Control: production of eprint (0) enabled
\begin{thebibliography}{40}%
\makeatletter
\providecommand \@ifxundefined [1]{%
 \@ifx{#1\undefined}
}%
\providecommand \@ifnum [1]{%
 \ifnum #1\expandafter \@firstoftwo
 \else \expandafter \@secondoftwo
 \fi
}%
\providecommand \@ifx [1]{%
 \ifx #1\expandafter \@firstoftwo
 \else \expandafter \@secondoftwo
 \fi
}%
\providecommand \natexlab [1]{#1}%
\providecommand \enquote  [1]{``#1''}%
\providecommand \bibnamefont  [1]{#1}%
\providecommand \bibfnamefont [1]{#1}%
\providecommand \citenamefont [1]{#1}%
\providecommand \href@noop [0]{\@secondoftwo}%
\providecommand \href [0]{\begingroup \@sanitize@url \@href}%
\providecommand \@href[1]{\@@startlink{#1}\@@href}%
\providecommand \@@href[1]{\endgroup#1\@@endlink}%
\providecommand \@sanitize@url [0]{\catcode `\\12\catcode `\$12\catcode
  `\&12\catcode `\#12\catcode `\^12\catcode `\_12\catcode `\%12\relax}%
\providecommand \@@startlink[1]{}%
\providecommand \@@endlink[0]{}%
\providecommand \url  [0]{\begingroup\@sanitize@url \@url }%
\providecommand \@url [1]{\endgroup\@href {#1}{\urlprefix }}%
\providecommand \urlprefix  [0]{URL }%
\providecommand \Eprint [0]{\href }%
\providecommand \doibase [0]{https://doi.org/}%
\providecommand \selectlanguage [0]{\@gobble}%
\providecommand \bibinfo  [0]{\@secondoftwo}%
\providecommand \bibfield  [0]{\@secondoftwo}%
\providecommand \translation [1]{[#1]}%
\providecommand \BibitemOpen [0]{}%
\providecommand \bibitemStop [0]{}%
\providecommand \bibitemNoStop [0]{.\EOS\space}%
\providecommand \EOS [0]{\spacefactor3000\relax}%
\providecommand \BibitemShut  [1]{\csname bibitem#1\endcsname}%
\let\auto@bib@innerbib\@empty
%</preamble>
\bibitem [{\citenamefont {Wheeler}(1953)}]{Wheeler1953}%
  \BibitemOpen
  \bibfield  {author} {\bibinfo {author} {\bibfnamefont {J.~A.}\ \bibnamefont
  {Wheeler}},\ }\href {https://doi.org/10.1103/PhysRev.92.812} {\bibfield
  {journal} {\bibinfo  {journal} {Phys. Rev.}\ }\textbf {\bibinfo {volume}
  {92}},\ \bibinfo {pages} {812} (\bibinfo {year} {1953})}\BibitemShut
  {NoStop}%
\bibitem [{\citenamefont {Engfer}\ \emph {et~al.}(1974)\citenamefont {Engfer},
  \citenamefont {Schneuwly}, \citenamefont {Vuilleumier}, \citenamefont
  {Walter},\ and\ \citenamefont {Zehnder}}]{Engfer1974}%
  \BibitemOpen
  \bibfield  {author} {\bibinfo {author} {\bibfnamefont {R.}~\bibnamefont
  {Engfer}}, \bibinfo {author} {\bibfnamefont {H.}~\bibnamefont {Schneuwly}},
  \bibinfo {author} {\bibfnamefont {J.}~\bibnamefont {Vuilleumier}}, \bibinfo
  {author} {\bibfnamefont {H.}~\bibnamefont {Walter}},\ and\ \bibinfo {author}
  {\bibfnamefont {A.}~\bibnamefont {Zehnder}},\ }\href
  {https://doi.org/https://doi.org/10.1016/S0092-640X(74)80003-3} {\bibfield
  {journal} {\bibinfo  {journal} {At. Data Nucl. Data Tables}\ }\textbf
  {\bibinfo {volume} {14}},\ \bibinfo {pages} {509} (\bibinfo {year}
  {1974})}\BibitemShut {NoStop}%
\bibitem [{\citenamefont {Borie}\ and\ \citenamefont
  {Rinker}(1982)}]{Borie1982}%
  \BibitemOpen
  \bibfield  {author} {\bibinfo {author} {\bibfnamefont {E.}~\bibnamefont
  {Borie}}\ and\ \bibinfo {author} {\bibfnamefont {G.~A.}\ \bibnamefont
  {Rinker}},\ }\href {https://doi.org/10.1103/RevModPhys.54.67} {\bibfield
  {journal} {\bibinfo  {journal} {Rev. Mod. Phys.}\ }\textbf {\bibinfo {volume}
  {54}},\ \bibinfo {pages} {67} (\bibinfo {year} {1982})}\BibitemShut {NoStop}%
\bibitem [{\citenamefont {B{\"{u}}ttgenbach}(1984)}]{Buttgenbach1984}%
  \BibitemOpen
  \bibfield  {author} {\bibinfo {author} {\bibfnamefont {S.}~\bibnamefont
  {B{\"{u}}ttgenbach}},\ }\href {https://doi.org/10.1007/BF02043319} {\bibfield
   {journal} {\bibinfo  {journal} {Hyperfine Interact}\ }\textbf {\bibinfo
  {volume} {20}},\ \bibinfo {pages} {1} (\bibinfo {year} {1984})}\BibitemShut
  {NoStop}%
\bibitem [{\citenamefont {Knecht}\ \emph {et~al.}(2020)\citenamefont {Knecht},
  \citenamefont {Skawran},\ and\ \citenamefont {Vogiatzi}}]{Knecht2020}%
  \BibitemOpen
  \bibfield  {author} {\bibinfo {author} {\bibfnamefont {A.}~\bibnamefont
  {Knecht}}, \bibinfo {author} {\bibfnamefont {A.}~\bibnamefont {Skawran}},\
  and\ \bibinfo {author} {\bibfnamefont {S.~M.}\ \bibnamefont {Vogiatzi}},\
  }\href {https://doi.org/10.1140/epjp/s13360-020-00777-y} {\bibfield
  {journal} {\bibinfo  {journal} {Eur. Phys. J. Plus}\ }\textbf {\bibinfo
  {volume} {135}},\ \bibinfo {pages} {777} (\bibinfo {year}
  {2020})}\BibitemShut {NoStop}%
\bibitem [{\citenamefont {Valuev}\ \emph {et~al.}(2022)\citenamefont {Valuev},
  \citenamefont {Col\`o}, \citenamefont {Roca-Maza}, \citenamefont {Keitel},\
  and\ \citenamefont {Oreshkina}}]{Valuev2022}%
  \BibitemOpen
  \bibfield  {author} {\bibinfo {author} {\bibfnamefont {I.~A.}\ \bibnamefont
  {Valuev}}, \bibinfo {author} {\bibfnamefont {G.}~\bibnamefont {Col\`o}},
  \bibinfo {author} {\bibfnamefont {X.}~\bibnamefont {Roca-Maza}}, \bibinfo
  {author} {\bibfnamefont {C.~H.}\ \bibnamefont {Keitel}},\ and\ \bibinfo
  {author} {\bibfnamefont {N.~S.}\ \bibnamefont {Oreshkina}},\ }\bibfield
  {title} {\bibinfo {title} {Evidence against nuclear polarization as source of
  fine-structure anomalies in muonic atoms},\ }\href
  {https://doi.org/10.1103/PhysRevLett.128.203001} {\bibfield  {journal}
  {\bibinfo  {journal} {Phys. Rev. Lett.}\ }\textbf {\bibinfo {volume} {128}},\
  \bibinfo {pages} {203001} (\bibinfo {year} {2022})}\BibitemShut {NoStop}%
\bibitem [{\citenamefont {Oreshkina}(2022)}]{Oreshkina2022}%
  \BibitemOpen
  \bibfield  {author} {\bibinfo {author} {\bibfnamefont {N.~S.}\ \bibnamefont
  {Oreshkina}},\ }\bibfield  {title} {\bibinfo {title} {Self-energy correction
  to the energy levels of heavy muonic atoms},\ }\href
  {https://doi.org/10.1103/PhysRevResearch.4.L042040} {\bibfield  {journal}
  {\bibinfo  {journal} {Phys. Rev. Res.}\ }\textbf {\bibinfo {volume} {4}},\
  \bibinfo {pages} {L042040} (\bibinfo {year} {2022})}\BibitemShut {NoStop}%
\bibitem [{\citenamefont {Sanamyan}\ \emph {et~al.}(2023)\citenamefont
  {Sanamyan}, \citenamefont {Roberts},\ and\ \citenamefont
  {Ginges}}]{Sanamyan2023}%
  \BibitemOpen
  \bibfield  {author} {\bibinfo {author} {\bibfnamefont {G.}~\bibnamefont
  {Sanamyan}}, \bibinfo {author} {\bibfnamefont {B.~M.}\ \bibnamefont
  {Roberts}},\ and\ \bibinfo {author} {\bibfnamefont {J.~S.~M.}\ \bibnamefont
  {Ginges}},\ }\bibfield  {title} {\bibinfo {title} {{Empirical determination
  of the Bohr-Weisskopf effect in cesium and improved tests of precision atomic
  theory in searches for new physics}},\ }\href
  {https://doi.org/10.1103/PhysRevLett.130.053001} {\bibfield  {journal}
  {\bibinfo  {journal} {Phys. Rev. Lett.}\ }\textbf {\bibinfo {volume} {130}},\
  \bibinfo {pages} {053001} (\bibinfo {year} {2023})}\BibitemShut {NoStop}%
\bibitem [{\citenamefont {Yerokhin}\ and\ \citenamefont
  {Oreshkina}(2023)}]{Yerokhin2023}%
  \BibitemOpen
  \bibfield  {author} {\bibinfo {author} {\bibfnamefont {V.~A.}\ \bibnamefont
  {Yerokhin}}\ and\ \bibinfo {author} {\bibfnamefont {N.~S.}\ \bibnamefont
  {Oreshkina}},\ }\bibfield  {title} {\bibinfo {title} {{QED calculations of
  the nuclear recoil effect in muonic atoms}},\ }\href
  {https://doi.org/10.1103/PhysRevA.108.052824} {\bibfield  {journal} {\bibinfo
   {journal} {Phys. Rev. A}\ }\textbf {\bibinfo {volume} {108}},\ \bibinfo
  {pages} {052824} (\bibinfo {year} {2023})}\BibitemShut {NoStop}%
\bibitem [{\citenamefont {Xie}\ \emph {et~al.}(2023)\citenamefont {Xie},
  \citenamefont {Naito}, \citenamefont {Li},\ and\ \citenamefont
  {Liang}}]{Xie2023}%
  \BibitemOpen
  \bibfield  {author} {\bibinfo {author} {\bibfnamefont {H.~H.}\ \bibnamefont
  {Xie}}, \bibinfo {author} {\bibfnamefont {T.}~\bibnamefont {Naito}}, \bibinfo
  {author} {\bibfnamefont {J.}~\bibnamefont {Li}},\ and\ \bibinfo {author}
  {\bibfnamefont {H.}~\bibnamefont {Liang}},\ }\bibfield  {title} {\bibinfo
  {title} {{Revisiting the extraction of charge radii of $^{40}\text{Ca}$ and
  $^{208}\text{Pb}$ with muonic atom spectroscopy}},\ }\href
  {https://www.scopus.com/inward/record.uri?eid=2-s2.0-85174055267&doi=10.1016%2fj.physletb.2023.138232&partnerID=40&md5=c414c23c2b4aeaddcf5556496681c8d5}
  {\bibfield  {journal} {\bibinfo  {journal} {Physics Letters, Section B:
  Nuclear, Elementary Particle and High-Energy Physics}\ }\textbf {\bibinfo
  {volume} {846}} (\bibinfo {year} {2023})}\BibitemShut {NoStop}%
\bibitem [{\citenamefont {Ginges}\ and\ \citenamefont
  {Flambaum}(2004)}]{GingesReview2004}%
  \BibitemOpen
  \bibfield  {author} {\bibinfo {author} {\bibfnamefont {J.~S.~M.}\
  \bibnamefont {Ginges}}\ and\ \bibinfo {author} {\bibfnamefont {V.~V.}\
  \bibnamefont {Flambaum}},\ }\href
  {https://doi.org/10.1016/j.physrep.2004.03.005} {\bibfield  {journal}
  {\bibinfo  {journal} {Phys. Rep.}\ }\textbf {\bibinfo {volume} {397}},\
  \bibinfo {pages} {63} (\bibinfo {year} {2004})}\BibitemShut {NoStop}%
\bibitem [{\citenamefont {Safronova}\ \emph {et~al.}(2018)\citenamefont
  {Safronova}, \citenamefont {Budker}, \citenamefont {DeMille}, \citenamefont
  {Jackson~Kimball}, \citenamefont {Derevianko},\ and\ \citenamefont
  {Clark}}]{SafronovaReview2018}%
  \BibitemOpen
  \bibfield  {author} {\bibinfo {author} {\bibfnamefont {M.~S.}\ \bibnamefont
  {Safronova}}, \bibinfo {author} {\bibfnamefont {D.}~\bibnamefont {Budker}},
  \bibinfo {author} {\bibfnamefont {D.}~\bibnamefont {DeMille}}, \bibinfo
  {author} {\bibfnamefont {D.~F.}\ \bibnamefont {Jackson~Kimball}}, \bibinfo
  {author} {\bibfnamefont {A.}~\bibnamefont {Derevianko}},\ and\ \bibinfo
  {author} {\bibfnamefont {C.~W.}\ \bibnamefont {Clark}},\ }\href
  {https://doi.org/10.1103/RevModPhys.90.025008} {\bibfield  {journal}
  {\bibinfo  {journal} {Rev. Mod. Phys.}\ }\textbf {\bibinfo {volume} {90}},\
  \bibinfo {pages} {025008} (\bibinfo {year} {2018})}\BibitemShut {NoStop}%
\bibitem [{\citenamefont {Antognini}\ \emph {et~al.}(2020)\citenamefont
  {Antognini}, \citenamefont {Berger}, \citenamefont {Cocolios}, \citenamefont
  {Dressler}, \citenamefont {Eichler}, \citenamefont {Eggenberger},
  \citenamefont {Indelicato}, \citenamefont {Jungmann}, \citenamefont {Keitel},
  \citenamefont {Kirch}, \citenamefont {Knecht}, \citenamefont {Michel},
  \citenamefont {Nuber}, \citenamefont {Oreshkina}, \citenamefont {Ouf},
  \citenamefont {Papa}, \citenamefont {Pohl}, \citenamefont {Pospelov},
  \citenamefont {Rapisarda}, \citenamefont {Ritjoho}, \citenamefont {Roccia},
  \citenamefont {Severijns}, \citenamefont {Skawran}, \citenamefont {Vogiatzi},
  \citenamefont {Wauters},\ and\ \citenamefont
  {Willmann}}]{PhysRevC.101.054313}%
  \BibitemOpen
  \bibfield  {author} {\bibinfo {author} {\bibfnamefont {A.}~\bibnamefont
  {Antognini}}, \bibinfo {author} {\bibfnamefont {N.}~\bibnamefont {Berger}},
  \bibinfo {author} {\bibfnamefont {T.~E.}\ \bibnamefont {Cocolios}}, \bibinfo
  {author} {\bibfnamefont {R.}~\bibnamefont {Dressler}}, \bibinfo {author}
  {\bibfnamefont {R.}~\bibnamefont {Eichler}}, \bibinfo {author} {\bibfnamefont
  {A.}~\bibnamefont {Eggenberger}}, \bibinfo {author} {\bibfnamefont
  {P.}~\bibnamefont {Indelicato}}, \bibinfo {author} {\bibfnamefont
  {K.}~\bibnamefont {Jungmann}}, \bibinfo {author} {\bibfnamefont {C.~H.}\
  \bibnamefont {Keitel}}, \bibinfo {author} {\bibfnamefont {K.}~\bibnamefont
  {Kirch}}, \bibinfo {author} {\bibfnamefont {A.}~\bibnamefont {Knecht}},
  \bibinfo {author} {\bibfnamefont {N.}~\bibnamefont {Michel}}, \bibinfo
  {author} {\bibfnamefont {J.}~\bibnamefont {Nuber}}, \bibinfo {author}
  {\bibfnamefont {N.~S.}\ \bibnamefont {Oreshkina}}, \bibinfo {author}
  {\bibfnamefont {A.}~\bibnamefont {Ouf}}, \bibinfo {author} {\bibfnamefont
  {A.}~\bibnamefont {Papa}}, \bibinfo {author} {\bibfnamefont {R.}~\bibnamefont
  {Pohl}}, \bibinfo {author} {\bibfnamefont {M.}~\bibnamefont {Pospelov}},
  \bibinfo {author} {\bibfnamefont {E.}~\bibnamefont {Rapisarda}}, \bibinfo
  {author} {\bibfnamefont {N.}~\bibnamefont {Ritjoho}}, \bibinfo {author}
  {\bibfnamefont {S.}~\bibnamefont {Roccia}}, \bibinfo {author} {\bibfnamefont
  {N.}~\bibnamefont {Severijns}}, \bibinfo {author} {\bibfnamefont
  {A.}~\bibnamefont {Skawran}}, \bibinfo {author} {\bibfnamefont {S.~M.}\
  \bibnamefont {Vogiatzi}}, \bibinfo {author} {\bibfnamefont {F.}~\bibnamefont
  {Wauters}},\ and\ \bibinfo {author} {\bibfnamefont {L.}~\bibnamefont
  {Willmann}},\ }\bibfield  {title} {\bibinfo {title} {Measurement of the
  quadrupole moment of $^{185}\mathrm{Re}$ and $^{187}\mathrm{Re}$ from the
  hyperfine structure of muonic x rays},\ }\href
  {https://doi.org/10.1103/PhysRevC.101.054313} {\bibfield  {journal} {\bibinfo
   {journal} {Phys. Rev. C}\ }\textbf {\bibinfo {volume} {101}},\ \bibinfo
  {pages} {054313} (\bibinfo {year} {2020})}\BibitemShut {NoStop}%
\bibitem [{\citenamefont {Plunien}\ \emph {et~al.}(1989)\citenamefont
  {Plunien}, \citenamefont {M\"uller}, \citenamefont {Greiner},\ and\
  \citenamefont {Soff}}]{Plunien1989}%
  \BibitemOpen
  \bibfield  {author} {\bibinfo {author} {\bibfnamefont {G.}~\bibnamefont
  {Plunien}}, \bibinfo {author} {\bibfnamefont {B.}~\bibnamefont {M\"uller}},
  \bibinfo {author} {\bibfnamefont {W.}~\bibnamefont {Greiner}},\ and\ \bibinfo
  {author} {\bibfnamefont {G.}~\bibnamefont {Soff}},\ }\bibfield  {title}
  {\bibinfo {title} {{Nuclear polarization contribution to the Lamb shift in
  heavy atoms}},\ }\href@noop {} {\bibfield  {journal} {\bibinfo  {journal}
  {Phys. Rev. A}\ }\textbf {\bibinfo {volume} {39}},\ \bibinfo {pages} {5428}
  (\bibinfo {year} {1989})}\BibitemShut {NoStop}%
\bibitem [{\citenamefont {Beier}(2000)}]{Beier2000}%
  \BibitemOpen
  \bibfield  {author} {\bibinfo {author} {\bibfnamefont {T.}~\bibnamefont
  {Beier}},\ }\href {https://doi.org/10.1016/S0370-1573(00)00071-5} {\bibfield
  {journal} {\bibinfo  {journal} {Phys. Rep.}\ }\textbf {\bibinfo {volume}
  {339}},\ \bibinfo {pages} {79} (\bibinfo {year} {2000})}\BibitemShut
  {NoStop}%
\bibitem [{\citenamefont {Tomalak}(2019)}]{Tomalak2019}%
  \BibitemOpen
  \bibfield  {author} {\bibinfo {author} {\bibfnamefont {O.}~\bibnamefont
  {Tomalak}},\ }\bibfield  {title} {\bibinfo {title} {Two-photon exchange on
  the neutron and the hyperfine splitting},\ }\href
  {https://doi.org/10.1103/physrevd.99.056018} {\bibfield  {journal} {\bibinfo
  {journal} {Physical Review D}\ }\textbf {\bibinfo {volume} {99}},\ \bibinfo
  {pages} {056018} (\bibinfo {year} {2019})}\BibitemShut {NoStop}%
\bibitem [{\citenamefont {Patk{\'{o}}{\v{s}}}\ \emph
  {et~al.}(2023)\citenamefont {Patk{\'{o}}{\v{s}}}, \citenamefont {Yerokhin},\
  and\ \citenamefont {Pachucki}}]{Patkos2023}%
  \BibitemOpen
  \bibfield  {author} {\bibinfo {author} {\bibfnamefont {V.}~\bibnamefont
  {Patk{\'{o}}{\v{s}}}}, \bibinfo {author} {\bibfnamefont {V.~A.}\ \bibnamefont
  {Yerokhin}},\ and\ \bibinfo {author} {\bibfnamefont {K.}~\bibnamefont
  {Pachucki}},\ }\bibfield  {title} {\bibinfo {title} {Nuclear polarizability
  effects in $^3${He}$^+$ hyperfine splitting},\ }\href
  {https://doi.org/10.1103/physreva.107.052802} {\bibfield  {journal} {\bibinfo
   {journal} {Physical Review A}\ }\textbf {\bibinfo {volume} {107}},\ \bibinfo
  {pages} {052802} (\bibinfo {year} {2023})}\BibitemShut {NoStop}%
\bibitem [{\citenamefont {Nefiodov}\ \emph {et~al.}(2003)\citenamefont
  {Nefiodov}, \citenamefont {Plunien},\ and\ \citenamefont
  {Soff}}]{Nefiodov2003}%
  \BibitemOpen
  \bibfield  {author} {\bibinfo {author} {\bibfnamefont {A.}~\bibnamefont
  {Nefiodov}}, \bibinfo {author} {\bibfnamefont {G.}~\bibnamefont {Plunien}},\
  and\ \bibinfo {author} {\bibfnamefont {G.}~\bibnamefont {Soff}},\ }\bibfield
  {title} {\bibinfo {title} {Nuclear-polarization effect to the hyperfine
  structure in heavy multicharged ions},\ }\href
  {https://doi.org/https://doi.org/10.1016/S0370-2693(02)03093-9} {\bibfield
  {journal} {\bibinfo  {journal} {Physics Letters B}\ }\textbf {\bibinfo
  {volume} {552}},\ \bibinfo {pages} {35} (\bibinfo {year} {2003})}\BibitemShut
  {NoStop}%
\bibitem [{\citenamefont {Volotka}\ and\ \citenamefont
  {Plunien}(2014)}]{Volotka2014}%
  \BibitemOpen
  \bibfield  {author} {\bibinfo {author} {\bibfnamefont {A.~V.}\ \bibnamefont
  {Volotka}}\ and\ \bibinfo {author} {\bibfnamefont {G.}~\bibnamefont
  {Plunien}},\ }\bibfield  {title} {\bibinfo {title} {{Nuclear Polarization
  Study: New Frontiers for Tests of QED in Heavy Highly Charged Ions}},\ }\href
  {https://doi.org/10.1103/PhysRevLett.113.023002} {\bibfield  {journal}
  {\bibinfo  {journal} {Phys. Rev. Lett.}\ }\textbf {\bibinfo {volume} {113}},\
  \bibinfo {pages} {023002} (\bibinfo {year} {2014})}\BibitemShut {NoStop}%
\bibitem [{\citenamefont {Angeli}\ and\ \citenamefont
  {Marinova}(2013)}]{Angeli2013}%
  \BibitemOpen
  \bibfield  {author} {\bibinfo {author} {\bibfnamefont {I.}~\bibnamefont
  {Angeli}}\ and\ \bibinfo {author} {\bibfnamefont {K.~P.}\ \bibnamefont
  {Marinova}},\ }\href
  {https://doi.org/https://doi.org/10.1016/j.adt.2011.12.006} {\bibfield
  {journal} {\bibinfo  {journal} {At. Data Nucl. Data Tables}\ }\textbf
  {\bibinfo {volume} {99}},\ \bibinfo {pages} {69} (\bibinfo {year}
  {2013})}\BibitemShut {NoStop}%
\bibitem [{\citenamefont {Volotka}\ \emph {et~al.}(2008)\citenamefont
  {Volotka}, \citenamefont {Glazov}, \citenamefont {Tupitsyn}, \citenamefont
  {Oreshkina}, \citenamefont {Plunien},\ and\ \citenamefont
  {Shabaev}}]{Volotka2008}%
  \BibitemOpen
  \bibfield  {author} {\bibinfo {author} {\bibfnamefont {A.~V.}\ \bibnamefont
  {Volotka}}, \bibinfo {author} {\bibfnamefont {D.~A.}\ \bibnamefont {Glazov}},
  \bibinfo {author} {\bibfnamefont {I.~I.}\ \bibnamefont {Tupitsyn}}, \bibinfo
  {author} {\bibfnamefont {N.~S.}\ \bibnamefont {Oreshkina}}, \bibinfo {author}
  {\bibfnamefont {G.}~\bibnamefont {Plunien}},\ and\ \bibinfo {author}
  {\bibfnamefont {V.~M.}\ \bibnamefont {Shabaev}},\ }\href
  {https://doi.org/10.1103/PhysRevA.78.062507} {\bibfield  {journal} {\bibinfo
  {journal} {Phys. Rev. A}\ }\textbf {\bibinfo {volume} {78}},\ \bibinfo
  {pages} {062507} (\bibinfo {year} {2008})}\BibitemShut {NoStop}%
\bibitem [{\citenamefont {Elizarov}\ \emph {et~al.}(2006)\citenamefont
  {Elizarov}, \citenamefont {Shabaev}, \citenamefont {Oreshkina}, \citenamefont
  {Tupitsyn},\ and\ \citenamefont {St{\"{o}}hlker}}]{Elizarov2006}%
  \BibitemOpen
  \bibfield  {author} {\bibinfo {author} {\bibfnamefont {A.~A.}\ \bibnamefont
  {Elizarov}}, \bibinfo {author} {\bibfnamefont {V.~M.}\ \bibnamefont
  {Shabaev}}, \bibinfo {author} {\bibfnamefont {N.~S.}\ \bibnamefont
  {Oreshkina}}, \bibinfo {author} {\bibfnamefont {I.~I.}\ \bibnamefont
  {Tupitsyn}},\ and\ \bibinfo {author} {\bibfnamefont {T.}~\bibnamefont
  {St{\"{o}}hlker}},\ }\href {https://doi.org/10.1134/S0030400X0603009X}
  {\bibfield  {journal} {\bibinfo  {journal} {Opt. Spectrosc.}\ }\textbf
  {\bibinfo {volume} {100}},\ \bibinfo {pages} {361} (\bibinfo {year}
  {2006})}\BibitemShut {NoStop}%
\bibitem [{\citenamefont {Michel}\ \emph {et~al.}(2017)\citenamefont {Michel},
  \citenamefont {Oreshkina},\ and\ \citenamefont {Keitel}}]{Michel2017}%
  \BibitemOpen
  \bibfield  {author} {\bibinfo {author} {\bibfnamefont {N.}~\bibnamefont
  {Michel}}, \bibinfo {author} {\bibfnamefont {N.~S.}\ \bibnamefont
  {Oreshkina}},\ and\ \bibinfo {author} {\bibfnamefont {C.~H.}\ \bibnamefont
  {Keitel}},\ }\href {https://doi.org/10.1103/PhysRevA.96.032510} {\bibfield
  {journal} {\bibinfo  {journal} {Phys. Rev. A}\ }\textbf {\bibinfo {volume}
  {96}},\ \bibinfo {pages} {032510} (\bibinfo {year} {2017})}\BibitemShut
  {NoStop}%
\bibitem [{\citenamefont {Yerokhin}\ and\ \citenamefont
  {Shabaev}(2015)}]{Yerokhin_Lamb_2011}%
  \BibitemOpen
  \bibfield  {author} {\bibinfo {author} {\bibfnamefont {V.~A.}\ \bibnamefont
  {Yerokhin}}\ and\ \bibinfo {author} {\bibfnamefont {V.~M.}\ \bibnamefont
  {Shabaev}},\ }\bibfield  {title} {\bibinfo {title} {{Lamb Shift of $n=1$ and
  $n=2$ States of Hydrogen-like Atoms, $1 \leq Z \leq 110$}},\ }\href
  {https://doi.org/10.1063/1.4927487} {\bibfield  {journal} {\bibinfo
  {journal} {{Journal of Physical and Chemical Reference Data}}\ }\textbf
  {\bibinfo {volume} {44}},\ \bibinfo {pages} {033103} (\bibinfo {year}
  {2015})}\BibitemShut {NoStop}%
\bibitem [{\citenamefont {Shabaev}\ \emph {et~al.}(1997)\citenamefont
  {Shabaev}, \citenamefont {Tomaselli}, \citenamefont {K{\"{u}}hl},
  \citenamefont {Artemyev},\ and\ \citenamefont {Yerokhin}}]{Shabaev1997}%
  \BibitemOpen
  \bibfield  {author} {\bibinfo {author} {\bibfnamefont {V.~M.}\ \bibnamefont
  {Shabaev}}, \bibinfo {author} {\bibfnamefont {M.}~\bibnamefont {Tomaselli}},
  \bibinfo {author} {\bibfnamefont {T.}~\bibnamefont {K{\"{u}}hl}}, \bibinfo
  {author} {\bibfnamefont {A.~N.}\ \bibnamefont {Artemyev}},\ and\ \bibinfo
  {author} {\bibfnamefont {V.~A.}\ \bibnamefont {Yerokhin}},\ }\href
  {https://doi.org/10.1103/PhysRevA.56.252} {\bibfield  {journal} {\bibinfo
  {journal} {Phys. Rev. A}\ }\textbf {\bibinfo {volume} {56}},\ \bibinfo
  {pages} {252} (\bibinfo {year} {1997})}\BibitemShut {NoStop}%
\bibitem [{\citenamefont {Artemyev}\ \emph {et~al.}(2001)\citenamefont
  {Artemyev}, \citenamefont {Shabaev}, \citenamefont {Plunien}, \citenamefont
  {Soff},\ and\ \citenamefont {Yerokhin}}]{Artemyev2001}%
  \BibitemOpen
  \bibfield  {author} {\bibinfo {author} {\bibfnamefont {A.~N.}\ \bibnamefont
  {Artemyev}}, \bibinfo {author} {\bibfnamefont {V.~M.}\ \bibnamefont
  {Shabaev}}, \bibinfo {author} {\bibfnamefont {G.}~\bibnamefont {Plunien}},
  \bibinfo {author} {\bibfnamefont {G.}~\bibnamefont {Soff}},\ and\ \bibinfo
  {author} {\bibfnamefont {V.~A.}\ \bibnamefont {Yerokhin}},\ }\href
  {https://doi.org/10.1103/PhysRevA.63.062504} {\bibfield  {journal} {\bibinfo
  {journal} {Phys. Rev. A}\ }\textbf {\bibinfo {volume} {63}},\ \bibinfo
  {pages} {062504} (\bibinfo {year} {2001})}\BibitemShut {NoStop}%
\bibitem [{\citenamefont {Shabaev}\ \emph {et~al.}(2001)\citenamefont
  {Shabaev}, \citenamefont {Artemyev}, \citenamefont {Yerokhin}, \citenamefont
  {Zherebtsov},\ and\ \citenamefont {Soff}}]{Shabaev2001}%
  \BibitemOpen
  \bibfield  {author} {\bibinfo {author} {\bibfnamefont {V.~M.}\ \bibnamefont
  {Shabaev}}, \bibinfo {author} {\bibfnamefont {A.~N.}\ \bibnamefont
  {Artemyev}}, \bibinfo {author} {\bibfnamefont {V.~A.}\ \bibnamefont
  {Yerokhin}}, \bibinfo {author} {\bibfnamefont {O.~M.}\ \bibnamefont
  {Zherebtsov}},\ and\ \bibinfo {author} {\bibfnamefont {G.}~\bibnamefont
  {Soff}},\ }\href {https://doi.org/10.1103/PhysRevLett.86.3959} {\bibfield
  {journal} {\bibinfo  {journal} {Phys. Rev. Lett.}\ }\textbf {\bibinfo
  {volume} {86}},\ \bibinfo {pages} {3959} (\bibinfo {year}
  {2001})}\BibitemShut {NoStop}%
\bibitem [{\citenamefont {Ginges}\ and\ \citenamefont
  {Volotka}(2018)}]{Ginges2018}%
  \BibitemOpen
  \bibfield  {author} {\bibinfo {author} {\bibfnamefont {J.~S.~M.}\
  \bibnamefont {Ginges}}\ and\ \bibinfo {author} {\bibfnamefont {A.~V.}\
  \bibnamefont {Volotka}},\ }\href {https://doi.org/10.1103/PhysRevA.98.032504}
  {\bibfield  {journal} {\bibinfo  {journal} {Phys. Rev. A}\ }\textbf {\bibinfo
  {volume} {98}},\ \bibinfo {pages} {032504} (\bibinfo {year}
  {2018})}\BibitemShut {NoStop}%
\bibitem [{\citenamefont {Skripnikov}(2020)}]{Skripnikov2020}%
  \BibitemOpen
  \bibfield  {author} {\bibinfo {author} {\bibfnamefont {L.~V.}\ \bibnamefont
  {Skripnikov}},\ }\href {https://doi.org/10.1063/5.0024103} {\bibfield
  {journal} {\bibinfo  {journal} {J. Chem. Phys.}\ }\textbf {\bibinfo {volume}
  {153}},\ \bibinfo {pages} {114114} (\bibinfo {year} {2020})}\BibitemShut
  {NoStop}%
\bibitem [{\citenamefont {Roberts}\ \emph {et~al.}(2022)\citenamefont
  {Roberts}, \citenamefont {Ranclaud},\ and\ \citenamefont
  {Ginges}}]{Roberts2021scr}%
  \BibitemOpen
  \bibfield  {author} {\bibinfo {author} {\bibfnamefont {B.~M.}\ \bibnamefont
  {Roberts}}, \bibinfo {author} {\bibfnamefont {P.~G.}\ \bibnamefont
  {Ranclaud}},\ and\ \bibinfo {author} {\bibfnamefont {J.~S.~M.}\ \bibnamefont
  {Ginges}},\ }\href {https://doi.org/10.1103/PhysRevA.105.052802} {\bibfield
  {journal} {\bibinfo  {journal} {Phys. Rev. A}\ }\textbf {\bibinfo {volume}
  {105}},\ \bibinfo {pages} {052802} (\bibinfo {year} {2022})}\BibitemShut
  {NoStop}%
\bibitem [{\citenamefont {Beiersdorfer}\ \emph {et~al.}(2001)\citenamefont
  {Beiersdorfer}, \citenamefont {Utter}, \citenamefont {Wong}, \citenamefont
  {{Crespo L{\'{o}}pez-Urrutia}}, \citenamefont {Britten}, \citenamefont
  {Chen}, \citenamefont {Harris}, \citenamefont {Thoe}, \citenamefont {Thorn},
  \citenamefont {Tr{\"{a}}bert}, \citenamefont {Gustavsson}, \citenamefont
  {Forss{\'{e}}n},\ and\ \citenamefont
  {M{\aa}rtensson-Pendrill}}]{Beiersdorfer2001}%
  \BibitemOpen
  \bibfield  {author} {\bibinfo {author} {\bibfnamefont {P.}~\bibnamefont
  {Beiersdorfer}}, \bibinfo {author} {\bibfnamefont {S.~B.}\ \bibnamefont
  {Utter}}, \bibinfo {author} {\bibfnamefont {K.~L.}\ \bibnamefont {Wong}},
  \bibinfo {author} {\bibfnamefont {J.~R.}\ \bibnamefont {{Crespo
  L{\'{o}}pez-Urrutia}}}, \bibinfo {author} {\bibfnamefont {J.~A.}\
  \bibnamefont {Britten}}, \bibinfo {author} {\bibfnamefont {H.}~\bibnamefont
  {Chen}}, \bibinfo {author} {\bibfnamefont {C.~L.}\ \bibnamefont {Harris}},
  \bibinfo {author} {\bibfnamefont {R.~S.}\ \bibnamefont {Thoe}}, \bibinfo
  {author} {\bibfnamefont {D.~B.}\ \bibnamefont {Thorn}}, \bibinfo {author}
  {\bibfnamefont {E.}~\bibnamefont {Tr{\"{a}}bert}}, \bibinfo {author}
  {\bibfnamefont {M.~G.~H.}\ \bibnamefont {Gustavsson}}, \bibinfo {author}
  {\bibfnamefont {C.}~\bibnamefont {Forss{\'{e}}n}},\ and\ \bibinfo {author}
  {\bibfnamefont {A.~M.}\ \bibnamefont {M{\aa}rtensson-Pendrill}},\ }\href
  {https://doi.org/10.1103/PhysRevA.64.032506} {\bibfield  {journal} {\bibinfo
  {journal} {Phys. Rev. A}\ }\textbf {\bibinfo {volume} {64}},\ \bibinfo
  {pages} {032506} (\bibinfo {year} {2001})}\BibitemShut {NoStop}%
\bibitem [{\citenamefont {Ullmann}\ \emph {et~al.}(2017)\citenamefont
  {Ullmann}, \citenamefont {Andelkovic}, \citenamefont {Brandau}, \citenamefont
  {Dax}, \citenamefont {Geithner}, \citenamefont {Geppert}, \citenamefont
  {Gorges}, \citenamefont {Hammen}, \citenamefont {Hannen}, \citenamefont
  {Kaufmann}, \citenamefont {K{\"{o}}nig}, \citenamefont {Litvinov},
  \citenamefont {Lochmann}, \citenamefont {Maa{\ss}}, \citenamefont {Meisner},
  \citenamefont {Murb{\"{o}}ck}, \citenamefont {S{\'{a}}nchez}, \citenamefont
  {Schmidt}, \citenamefont {Schmidt}, \citenamefont {Steck}, \citenamefont
  {St{\"{o}}hlker}, \citenamefont {Thompson}, \citenamefont {Trageser},
  \citenamefont {Vollbrecht}, \citenamefont {Weinheimer},\ and\ \citenamefont
  {N{\"{o}}rtersha{\"{u}}ser}}]{Ullmann2017}%
  \BibitemOpen
  \bibfield  {author} {\bibinfo {author} {\bibfnamefont {J.}~\bibnamefont
  {Ullmann}}, \bibinfo {author} {\bibfnamefont {Z.}~\bibnamefont {Andelkovic}},
  \bibinfo {author} {\bibfnamefont {C.}~\bibnamefont {Brandau}}, \bibinfo
  {author} {\bibfnamefont {A.}~\bibnamefont {Dax}}, \bibinfo {author}
  {\bibfnamefont {W.}~\bibnamefont {Geithner}}, \bibinfo {author}
  {\bibfnamefont {C.}~\bibnamefont {Geppert}}, \bibinfo {author} {\bibfnamefont
  {C.}~\bibnamefont {Gorges}}, \bibinfo {author} {\bibfnamefont
  {M.}~\bibnamefont {Hammen}}, \bibinfo {author} {\bibfnamefont
  {V.}~\bibnamefont {Hannen}}, \bibinfo {author} {\bibfnamefont
  {S.}~\bibnamefont {Kaufmann}}, \bibinfo {author} {\bibfnamefont
  {K.}~\bibnamefont {K{\"{o}}nig}}, \bibinfo {author} {\bibfnamefont {Y.~A.}\
  \bibnamefont {Litvinov}}, \bibinfo {author} {\bibfnamefont {M.}~\bibnamefont
  {Lochmann}}, \bibinfo {author} {\bibfnamefont {B.}~\bibnamefont {Maa{\ss}}},
  \bibinfo {author} {\bibfnamefont {J.}~\bibnamefont {Meisner}}, \bibinfo
  {author} {\bibfnamefont {T.}~\bibnamefont {Murb{\"{o}}ck}}, \bibinfo {author}
  {\bibfnamefont {R.}~\bibnamefont {S{\'{a}}nchez}}, \bibinfo {author}
  {\bibfnamefont {M.}~\bibnamefont {Schmidt}}, \bibinfo {author} {\bibfnamefont
  {S.}~\bibnamefont {Schmidt}}, \bibinfo {author} {\bibfnamefont
  {M.}~\bibnamefont {Steck}}, \bibinfo {author} {\bibfnamefont
  {T.}~\bibnamefont {St{\"{o}}hlker}}, \bibinfo {author} {\bibfnamefont
  {R.~C.}\ \bibnamefont {Thompson}}, \bibinfo {author} {\bibfnamefont
  {C.}~\bibnamefont {Trageser}}, \bibinfo {author} {\bibfnamefont
  {J.}~\bibnamefont {Vollbrecht}}, \bibinfo {author} {\bibfnamefont
  {C.}~\bibnamefont {Weinheimer}},\ and\ \bibinfo {author} {\bibfnamefont
  {W.}~\bibnamefont {N{\"{o}}rtersha{\"{u}}ser}},\ }\href
  {https://doi.org/10.1038/ncomms15484} {\bibfield  {journal} {\bibinfo
  {journal} {Nat. Commun.}\ }\textbf {\bibinfo {volume} {8}},\ \bibinfo {pages}
  {15484} (\bibinfo {year} {2017})}\BibitemShut {NoStop}%
\bibitem [{\citenamefont {Labzowsky}\ and\ \citenamefont
  {Nefiodov}(1994)}]{Labzowsky1994}%
  \BibitemOpen
  \bibfield  {author} {\bibinfo {author} {\bibfnamefont {L.}~\bibnamefont
  {Labzowsky}}\ and\ \bibinfo {author} {\bibfnamefont {A.}~\bibnamefont
  {Nefiodov}},\ }\bibfield  {title} {\bibinfo {title} {Analytic evaluation of
  the nuclear polarization contribution to the energy shift in heavy ions},\
  }\href {https://doi.org/https://doi.org/10.1016/0375-9601(94)90478-2}
  {\bibfield  {journal} {\bibinfo  {journal} {Phys. Lett. A}\ }\textbf
  {\bibinfo {volume} {188}},\ \bibinfo {pages} {371 } (\bibinfo {year}
  {1994})}\BibitemShut {NoStop}%
\bibitem [{\citenamefont {Nefiodov}\ \emph {et~al.}(1996)\citenamefont
  {Nefiodov}, \citenamefont {Labzowsky}, \citenamefont {Plunien},\ and\
  \citenamefont {Soff}}]{Nefiodov1996}%
  \BibitemOpen
  \bibfield  {author} {\bibinfo {author} {\bibfnamefont {A.}~\bibnamefont
  {Nefiodov}}, \bibinfo {author} {\bibfnamefont {L.}~\bibnamefont {Labzowsky}},
  \bibinfo {author} {\bibfnamefont {G.}~\bibnamefont {Plunien}},\ and\ \bibinfo
  {author} {\bibfnamefont {G.}~\bibnamefont {Soff}},\ }\bibfield  {title}
  {\bibinfo {title} {Nuclear polarization effects in spectra of multicharged
  ions},\ }\href {https://doi.org/10.1016/0375-9601(96)00650-0} {\bibfield
  {journal} {\bibinfo  {journal} {Phys. Lett. A}\ }\textbf {\bibinfo {volume}
  {222}},\ \bibinfo {pages} {227} (\bibinfo {year} {1996})}\BibitemShut
  {NoStop}%
\bibitem [{\citenamefont {Cakir}\ \emph {et~al.}(2020)\citenamefont {Cakir},
  \citenamefont {Oreshkina}, \citenamefont {Valuev}, \citenamefont {Debierre},
  \citenamefont {Yerokhin}, \citenamefont {Keitel},\ and\ \citenamefont
  {Harman}}]{Cakir2020}%
  \BibitemOpen
  \bibfield  {author} {\bibinfo {author} {\bibfnamefont {H.}~\bibnamefont
  {Cakir}}, \bibinfo {author} {\bibfnamefont {N.~S.}\ \bibnamefont
  {Oreshkina}}, \bibinfo {author} {\bibfnamefont {I.~A.}\ \bibnamefont
  {Valuev}}, \bibinfo {author} {\bibfnamefont {V.}~\bibnamefont {Debierre}},
  \bibinfo {author} {\bibfnamefont {V.~A.}\ \bibnamefont {Yerokhin}}, \bibinfo
  {author} {\bibfnamefont {C.~H.}\ \bibnamefont {Keitel}},\ and\ \bibinfo
  {author} {\bibfnamefont {Z.}~\bibnamefont {Harman}},\ }\href@noop {}
  {\bibinfo {title} {Improved access to the fine-structure constant with the
  simplest atomic systems}} (\bibinfo {year} {2020}),\ \Eprint
  {https://arxiv.org/abs/2006.14261} {arXiv:2006.14261 [physics.atom-ph]}
  \BibitemShut {NoStop}%
\bibitem [{\citenamefont {Schneider}\ \emph {et~al.}(2022)\citenamefont
  {Schneider}, \citenamefont {Sikora}, \citenamefont {Dickopf}, \citenamefont
  {M\"{u}ller}, \citenamefont {Oreshkina}, \citenamefont {Rischka},
  \citenamefont {Valuev}, \citenamefont {Ulmer}, \citenamefont {Walz},
  \citenamefont {Harman}, \citenamefont {Keitel}, \citenamefont {Mooser},\ and\
  \citenamefont {Blaum}}]{Schneider_2022}%
  \BibitemOpen
  \bibfield  {author} {\bibinfo {author} {\bibfnamefont {A.}~\bibnamefont
  {Schneider}}, \bibinfo {author} {\bibfnamefont {B.}~\bibnamefont {Sikora}},
  \bibinfo {author} {\bibfnamefont {S.}~\bibnamefont {Dickopf}}, \bibinfo
  {author} {\bibfnamefont {M.}~\bibnamefont {M\"{u}ller}}, \bibinfo {author}
  {\bibfnamefont {N.~S.}\ \bibnamefont {Oreshkina}}, \bibinfo {author}
  {\bibfnamefont {A.}~\bibnamefont {Rischka}}, \bibinfo {author} {\bibfnamefont
  {I.~A.}\ \bibnamefont {Valuev}}, \bibinfo {author} {\bibfnamefont
  {S.}~\bibnamefont {Ulmer}}, \bibinfo {author} {\bibfnamefont
  {J.}~\bibnamefont {Walz}}, \bibinfo {author} {\bibfnamefont {Z.}~\bibnamefont
  {Harman}}, \bibinfo {author} {\bibfnamefont {C.~H.}\ \bibnamefont {Keitel}},
  \bibinfo {author} {\bibfnamefont {A.}~\bibnamefont {Mooser}},\ and\ \bibinfo
  {author} {\bibfnamefont {K.}~\bibnamefont {Blaum}},\ }\bibfield  {title}
  {\bibinfo {title} {Direct measurement of the ${}^3${He}$^+$ magnetic
  moments},\ }\href {https://doi.org/10.1038/s41586-022-04761-7} {\bibfield
  {journal} {\bibinfo  {journal} {Nature}\ }\textbf {\bibinfo {volume} {606}},\
  \bibinfo {pages} {878} (\bibinfo {year} {2022})}\BibitemShut {NoStop}%
\bibitem [{\citenamefont {Kondev}(2021)}]{KONDEV2021509}%
  \BibitemOpen
  \bibfield  {author} {\bibinfo {author} {\bibfnamefont {F.}~\bibnamefont
  {Kondev}},\ }\bibfield  {title} {\bibinfo {title} {Nuclear data sheets for
  {A=203}},\ }\href {https://doi.org/https://doi.org/10.1016/j.nds.2021.09.002}
  {\bibfield  {journal} {\bibinfo  {journal} {Nuclear Data Sheets}\ }\textbf
  {\bibinfo {volume} {177}},\ \bibinfo {pages} {509} (\bibinfo {year}
  {2021})}\BibitemShut {NoStop}%
\bibitem [{\citenamefont {Kondev}(2020)}]{KONDEV20201}%
  \BibitemOpen
  \bibfield  {author} {\bibinfo {author} {\bibfnamefont {F.}~\bibnamefont
  {Kondev}},\ }\bibfield  {title} {\bibinfo {title} {Nuclear data sheets for
  {A=205}},\ }\href {https://doi.org/https://doi.org/10.1016/j.nds.2020.05.001}
  {\bibfield  {journal} {\bibinfo  {journal} {Nuclear Data Sheets}\ }\textbf
  {\bibinfo {volume} {166}},\ \bibinfo {pages} {1} (\bibinfo {year}
  {2020})}\BibitemShut {NoStop}%
\bibitem [{\citenamefont {Chen}\ and\ \citenamefont
  {Kondev}(2015)}]{CHEN2015373}%
  \BibitemOpen
  \bibfield  {author} {\bibinfo {author} {\bibfnamefont {J.}~\bibnamefont
  {Chen}}\ and\ \bibinfo {author} {\bibfnamefont {F.}~\bibnamefont {Kondev}},\
  }\bibfield  {title} {\bibinfo {title} {Nuclear data sheets for {A=209}},\
  }\href {https://doi.org/https://doi.org/10.1016/j.nds.2015.05.003} {\bibfield
   {journal} {\bibinfo  {journal} {Nuclear Data Sheets}\ }\textbf {\bibinfo
  {volume} {126}},\ \bibinfo {pages} {373} (\bibinfo {year}
  {2015})}\BibitemShut {NoStop}%
\bibitem [{\citenamefont {Valuev}\ and\ \citenamefont
  {Oreshkina}(2024)}]{Valuev2024}%
  \BibitemOpen
  \bibfield  {author} {\bibinfo {author} {\bibfnamefont {I.~A.}\ \bibnamefont
  {Valuev}}\ and\ \bibinfo {author} {\bibfnamefont {N.~S.}\ \bibnamefont
  {Oreshkina}},\ }\bibfield  {title} {\bibinfo {title} {Full leading-order
  nuclear polarization in highly charged ions},\ }\href
  {https://doi.org/10.1103/PhysRevA.109.042811} {\bibfield  {journal} {\bibinfo
   {journal} {Phys. Rev. A}\ }\textbf {\bibinfo {volume} {109}},\ \bibinfo
  {pages} {042811} (\bibinfo {year} {2024})}\BibitemShut {NoStop}%
\end{thebibliography}%

\end{document}